# Fe Nanoclusters on the Ge(001) Surface Studied by Scanning Tunneling Microscopy, Density Functional Theory Calculations and X-Ray Magnetic Circular Dichroism.

O. Lübben[1,*], S. A. Krasnikov[1], A. B. Preobrajenski[2], B. E. Murphy[1], and I. V. Shvets[1]

[1] Centre for Research on Adaptive Nanostructures and Nanodevices (CRANN), School of Physics, Trinity College Dublin, Dublin 2, Ireland;
[2] MAX-lab, Lund University, Box 118, 22100 Lund, Sweden.

**Abstract**
The growth of Fe nanoclusters on the Ge(001) surface has been studied using low-temperature scanning tunnelling microscopy (STM) and density functional theory (DFT) calculations. STM results indicate that Fe nucleates on the Ge(001) surface, forming well-ordered nanoclusters of uniform size. Depending on the preparation conditions, two types of nanoclusters were observed having either four or sixteen Fe atoms within a nanocluster. The results were confirmed by DFT calculations. Annealing the nanoclusters at 420 K leads to the formation of nanorow structures, due to cluster mobility at such temperature. The Fe nanoclusters and nanorow structures formed on the Ge(001) surface show a superparamagnetic behaviour as measured by X-ray magnetic circular dichroism.

**1. Introduction**

The self-assembly of atoms or molecules into ordered surface-supported nanostructures is one of the key topics in solid state physics and surface science [1–9]. One major reason for this attention is the prospect of controlling atomic scale structures on surfaces, which can lead to mass fabrication of usable systems and novel devices. A promising approach towards the control of self-assembly is the use of preformed surface templates onto which particular nanostructures can be arranged in a well-ordered fashion [10–12]. Semiconductor surfaces such as the Ge(001)-(2×1) reconstructed surface are suitable templates for the growth of well-ordered, uniformly-sized metal nanoclusters [13, 14]. Their size, shape, and the spacing between clusters are dictated by the substrate. Such metal-nanocluster – semiconductor systems can be considered to be examples of low-dimensional dilute magnetic semiconductors (DMS), as the interaction between transition metal clusters takes place via the semiconducting substrate [15]. Furthermore, the layout of clusters on the surface into a regular exposed system allows for good control of the separation between clusters, which cannot be achieved in most DMS three-dimensional systems.

DMS are promising materials for use in many technological applications and their study is important for future developments in nanotechnology as well as from the fundamental point of view. Such systems have attracted much attention recently since they can be utilized as essential building blocks in the field of spin-dependent electronic (spintronic) devices, providing a link between magnetism and semiconductor technologies [16–19]. The incorporation of ferromagnetic elements into semiconductor devices may lead to data processing and magnetic storage on a single chip. One of the most suitable host semiconductor materials for developing DMS is the group IV semiconductor germanium (band gap = 0.62 eV at 300 K [20]), that can be readily doped with magnetic elements [17, 21–24]. It was shown that germanium crystals doped with Cr and Fe exhibit ferromagnetic behaviour at 126 K and 233 K, respectively [24, 25], provided the concentration of dopants is high enough. Furthermore, the possibility of using ferromagnetic metals as a source of spin polarised electrons for injection into semiconductors has led to a strong interest in the growth of ferromagnetic layers on Ge [26–30].

---

[*] Author to whom any correspondence should be addressed. E-mail: luebbeno@tcd.ie

In the present work we use the Ge(001) surface as a template for the growth of ordered arrays of Fe nanoclusters. We employ scanning tunnelling microscopy (STM), X-ray magnetic circular dichroism (XMCD) and density functional theory (DFT) calculations to study the nucleation, structure, mobility and magnetic properties of the Fe nanoclusters. The results obtained provide important information on Fe nanoclusters grown on the Ge(001) surface and will be of value for the development of spin-electronics.

## 2. Experimental details

The STM experiments were performed at liquid nitrogen temperature (78 K), using a commercial instrument from Createc, in an ultra-high-vacuum (UHV) system consisting of an analysis chamber (with a base pressure of $2 \times 10^{-11}$ mbar) and a preparation chamber ($5 \times 10^{-11}$ mbar). An electrochemically-etched monocrystalline W(100) tip [31] was used to record STM images in constant current mode. The voltage $V_b$ corresponds to the sample bias with respect to the tip. No drift corrections have been applied to any of the STM images presented in this paper.

The germanium samples were cleaved from a 3 inch n-type (Sb doped, 25 $\Omega^{-1}$ cm$^{-1}$) (001)-oriented wafer (MaTeck). The Ge(001) surface was cleaned by cycles of Ne$^+$ ion sputtering at an energy of 0.6 keV and annealing at 925 K for 40 min. Sample temperature was measured using an optical pyrometer (Ircon UX20P). After the final anneal the sample was cooled down to 675 K at a rate of 10 K per minute to acquire the (2×1) surface reconstruction present at room temperature [32]. The substrate cleanliness was verified by low-energy electron diffraction (LEED) and STM. Fe was deposited from an electron-beam evaporator at a rate of 0.1 monolayer (ML) per minute. The substrate was kept at room temperature (RT) during deposition. After deposition the sample was transferred into the STM and cooled down to 78 K.

XMCD measurements were performed at the D1011 beamline at MAX-lab synchrotron in Lund, Sweden. Measurements were conducted at RT as well as at 150 K. Fe 2p X-ray absorption (XA) spectra were recorded using sample drain current. The spectra were normalised to the background curves measured from a clean substrate. The photon energy resolution was set to 200 meV at the Fe $L_3$-edge (~710 eV). For XMCD measurements a switchable magnetic field of 0.05 T was applied.

## 3. Results and discussion

The Ge(001)-(2×1) surface reconstruction is obtained through the pairing of nearest-neighbour surface atoms, forming dimer rows oriented along the [1–11] crystallographic direction [32]. The c(4×2) reconstruction is observed on the Ge(001) surface at temperatures below 200 K [33]. At RT, weak quarter-order LEED spots have occasionally been observed on the surface, consistent with some localised 4×2 reconstruction [33]. The intensity of these quarter-order spots has been shown to increase with decreasing temperature, until at 200 K only the c(4×2) mesh is observed. This gradual change suggests that the (2×1) to c(4×2) two-dimensional phase transition is of second order [33]. The flip-flopping dimers of the (2×1) reconstruction either pass through a symmetric state or charge is transferred between the two dangling bonds of the dimer during their flip–flop motion. This results in a conductivity band for the (2×1) domains that is not present for the c(4×2) domains [34–36]. The buckled dimer forms the basis of the c(4×2) reconstruction. Unlike the (2×1) reconstruction, the dimers of the c(4×2) reconstruction are buckled out-of-phase (or anti-correlated) along the [110] oriented rows [33]. Furthermore, they do not flip-flop between the two possible configurations of the dimer, as in the case of the (2×1) reconstruction, but instead are frozen in one buckled configuration.

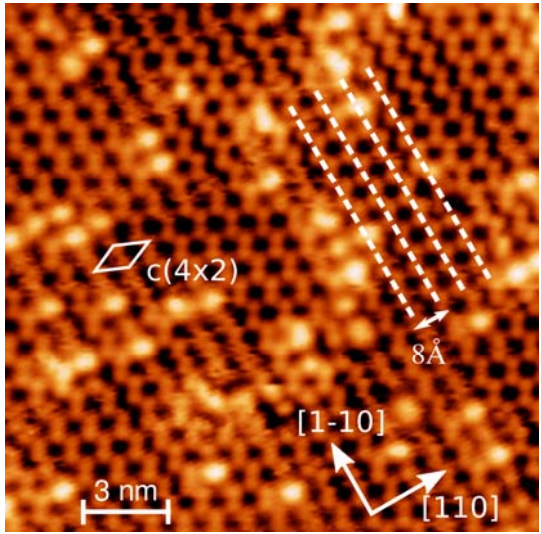

**Figure 1** Low-temperature STM image of 0.05 ML of Fe on the Ge(001) surface: $V_b = -1.7$ V, $I_t = 0.60$ nA, size 17.0 nm × 17.0 nm, 78 K. The unit cell of the underlying Ge(001)-c(4×2) reconstructed surface is shown in white.

If deposited at RT and immediately cooled down to 78 K, Fe forms small nanoclusters on the Ge(001) surface at low coverage (0.05–0.2 ML). A typical low temperature STM image of 0.05 ML of Fe on the Ge(001) surface is presented in Fig. 1, where bright protrusions correspond to Fe nanoclusters. All nanoclusters have a square shape and the uniform size of 4.5 Å ± 0.5 Å, which corresponds to four Fe atoms per cluster. These nanoclusters have an apparent corrugation height of 1.1 Å ± 0.1 Å with respect to the underlying substrate, which does not depend on the bias voltage.

It is clear that these two-dimensional nanoclusters follow the substrate dimer rows, although the separation between the clusters varies throughout the image. During the initial nucleation of the Fe nanoclusters, the interaction between Fe and the Ge(001) surface is strong enough to provide well-defined nucleation sites. Following this, additional Fe atoms diffuse on the surface to form nanoclusters around the initial nucleation sites. This indicates quite a strong Fe–Fe interaction. The underlying Ge(001)-c(4×2) reconstructed surface is also visible (Fig. 1). It exhibits large terraces with the Ge dimers forming zig-zag rows oriented along the [1–10] direction, in agreement with previous studies [10, 37]. These are indicated by white dashed lines in Fig. 1. The orientation of the dimer rows rotates by 90° between adjacent terraces. The separation of two adjacent dimers is 8 Å and 4 Å along the [110] and [1–10] directions, respectively.

If the sample is left at room temperature under UHV conditions for more than one hour after Fe deposition, larger clusters are observed on the Ge(001) surface, similar to results obtained by Jordan *et al.* [13]. Typical low temperature STM images of 0.2 ML of Fe on the Ge(001) surface are shown in Fig. 2. Under such preparation conditions Fe atoms form nanoclusters with a square shape with sides 9.0 Å ± 0.5 Å. Figs. 2a and 2b show occupied and unoccupied states of the Fe/Ge(001) system, respectively. The Ge(001)-c(4x2) reconstructed surface exhibits the charge distribution consistent with published reports for both occupied and unoccupied states [37]. In turn, the occupied and unoccupied states of Fe nanoclusters do not show a significant difference, which is typical for metals.

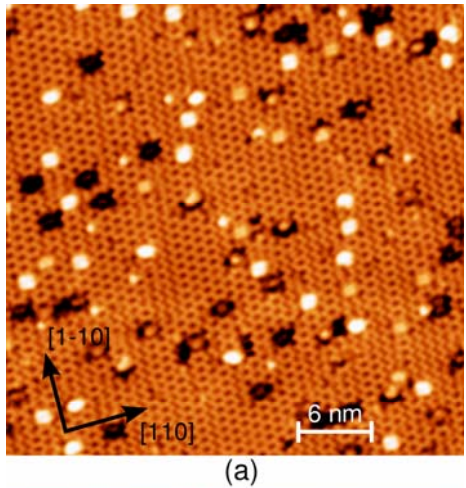
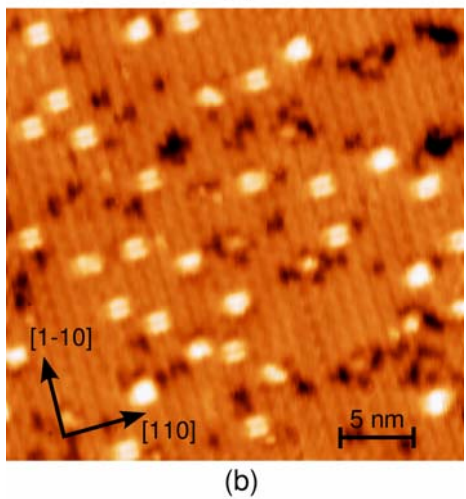

**Figure 2** Low-temperature STM images of 0.2 ML of Fe on the Ge(001) surface. The sample was left at RT under UHV conditions for more than one hour after Fe deposition. (a) STM image of occupied states: $V_b = -1.5$ V, $I_t = 0.50$ nA, size 30.0 nm × 30.0 nm, 78 K. (b) STM image of unoccupied states: $V_b = +1.5$ V, $I_t = 1.60$ nA, size 24.0 nm × 24.0 nm, 78 K.

It is clearly seen from Fig. 2 that the Fe nanoclusters exhibit an inner structure consisting of four smaller clusters, similar to those observed in Fig. 1. Furthermore, they have a size (9.0 Å × 9.0 Å), which is four times the size of smaller clusters (4.5 Å × 4.5 Å). These findings suggest that the large Fe nanoclusters are formed due to the migration of the smaller clusters or single Fe atoms at RT. The larger nanoclusters consist of sixteen Fe atoms. However, the apparent height of these nanoclusters is 1.1 Å ± 0.1 Å, indicating that they are also a single monolayer in height. The observation of these larger clusters after allowing the Fe/Ge(001) system to relax suggests that a cluster size of 9.0 Å × 9.0 Å is the energetically most favourable at room temperate. Furthermore, the self-assembly of Fe atoms into nanoclusters of a limited size at RT indicates that the strength of Fe–Fe interaction depends significantly on the size of the Fe nanocluster.

In order to confirm the number of Fe atoms in the nanoclusters formed on the Ge(001) surface, DFT calculations were performed using the Vienna *Ab-initio* Simulation Package (VASP) program. VASP implements a projected augmented waves basis set [38] and periodic boundary conditions. The electron exchange and correlation was simulated by local density approximation (LDA) pseudopotentials with a Ceperley-Alder exchange-correlation density functional [39]. The Ge(001) surface was simulated by periodic supercells formed by slabs consisting of six Ge unit layers and a vacuum slab of 15 Å [40]. The positions of the atoms in the layer most distant from the surface were constrained to simulate the bulk. A (6×3×1) k-point grid was used to sample the Brillouin zone. The applied energy cutoff was 215 eV. The Ge system has been initially relaxed to obtain the c(4×2) reconstruction of the

Ge(001) surface. The resulting c(4×2) reconstructed Ge(001) has been used as a "substrate" to model the Fe nanocluster.

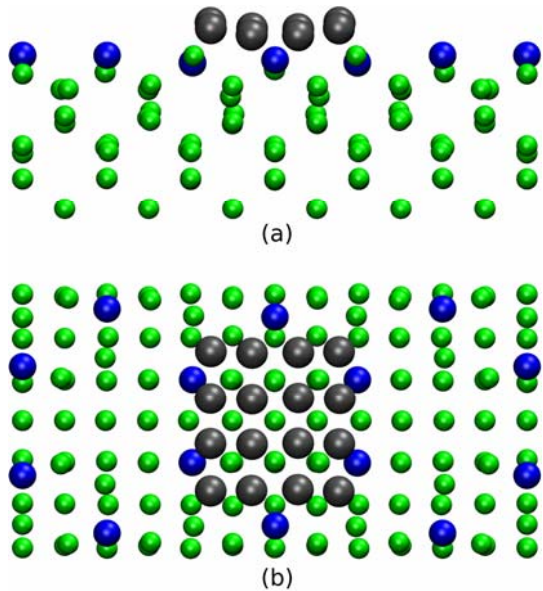

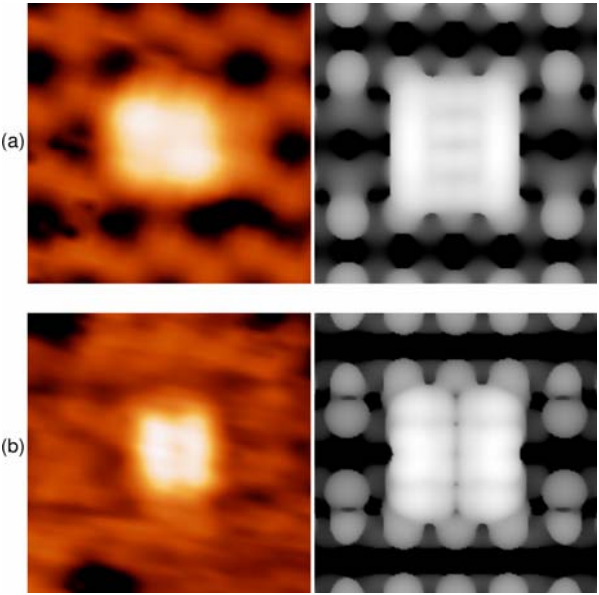

**Figure 3** Side (a) and top (b) views of the calculated relaxed structure of the Fe nanocluster, consisting of 16 Fe atoms, on the Ge(001)-c(4x2) reconstructed surface. The Fe atoms are denoted by large gray spheres, the Ge atoms of the surface layer by blue spheres, and the bulk Ge atoms by green spheres.

**Figure 4** (a) Occupied state STM image of the Fe nanocluster (3.0 nm × 3.0 nm, $V_b$ = –1.5 V), left panel, compared to the simulated partial charge density (from –1.5 V to $E_F$), right panel. (b) Unoccupied state STM image of the Fe nanocluster (4.0 nm × 4.0 nm, $V_b$ = +1.5 V), left panel, compared to the simulated partial charge density (from $E_F$ to +1.5 V), right panel.

For the simulation of the Fe/Ge(001) system an energy cutoff of 287 eV has been applied. The Γ-point has been used to sample the Brilliouin zone. The Fe/Ge(001) system was relaxed prior to any further calculations. Fe nanoclusters with 9, 12, 16, 17 and 18 atoms have been considered in order to fit an experimentally observed cluster size of 9.0 Å × 9.0 Å. Simulations of 12, 17 and 18 atoms have shown that some Fe atoms diffuse into the Ge substrate. This suggests bonding between Fe and Ge, which was ruled out on the basis of the Fe XA spectra, which have a shape typical of the XA spectrum of a metal. The system with 9 atoms resulted in a smaller cluster size than experimentally observed.

The Fe cluster with 16 atoms was found to have the lowest energy and a size of 9.0 Å × 9.0 Å, which is in excellent agreement with the experimental data. Six different starting configurations (geometrical positions) for the Fe nanocluster consisting of 16 atoms have been examined. Five of them either did not resemble the experimentally observed geometrical structure or had a higher energy than the simulated Fe nanocluster shown in Figs. 3 and 4. The resulting model of the Fe nanocluster on the Ge(001) surface is shown in Fig. 3. In order to further compare DFT results with the STM images, the partial charge density of the Fe on the Ge(001) surface has been simulated in the range between –1.5 V and the Fermi energy ($E_F$), as well as between $E_F$ and +1.5 V. The calculated images are compared with the STM data in Fig. 4 and show very good agreement.

In order to study the magnetic properties of the Fe nanoclusters, XMCD measurements were carried out. The X-ray absorption (XA) spectra from 0.5 ML of Fe on the Ge(001) surface shown in Fig. 5 were measured both at RT and at 150 K with the magnetic field of 0.05 T applied in two opposite directions. The relative intensities of the spectra are normalized to the same continuum jump at the photon energy of 745 eV, after subtraction of the background measured from a clean substrate. Nanoclusters, consisting of sixteen Fe atoms were grown on the sample. The XA spectra have a shape typical of the XA spectrum of a metal, indicating that no significant intermixing occurs at the Fe/Ge(001) interface, which is in agreement with previous studies [13, 30].

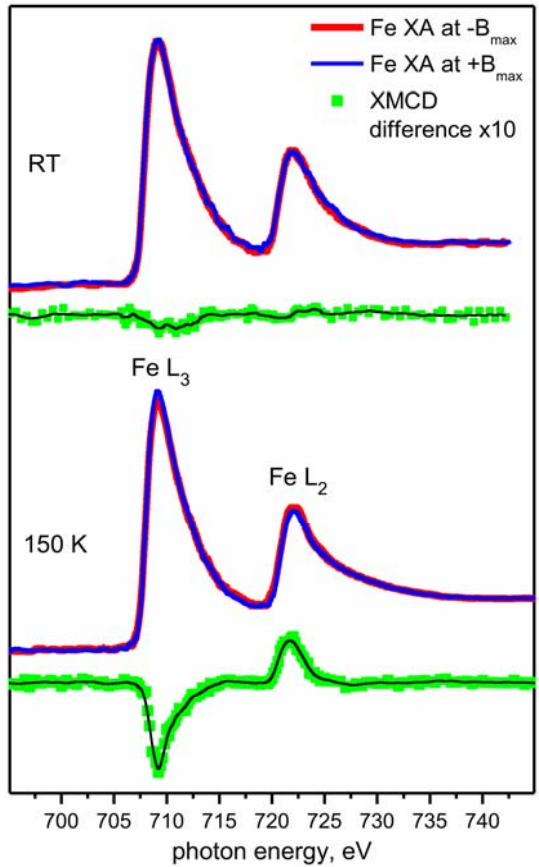

**Figure 5** Fe 2p XA spectra measured from 0.5 ML of Fe on the Ge(001) surface at RT (top) and at 150 K (bottom) with the magnetic field of 0.05 T applied in two opposite directions ($-B_{max}$ and $+B_{max}$). The relative intensities of the spectra have been normalized to the same continuum jump at the photon energy of 745 eV, after subtraction of the background measured from a clean substrate. The XMCD spectra are multiplied by a factor of 10 for clarity.

The XMCD spectra shown in Fig. 5 result from the subtraction of the XA spectrum taken at the maximum magnetic field ($+B_{max}$) applied in one direction (Fig. 5, blue curve) from the other spectrum taken at the maximum magnetic field ($-B_{max}$) applied in the opposite direction (Fig. 5, red curve). The XMCD spectrum taken at 150 K shows a prominent difference between the two opposite magnetic directions. In contrast, XMCD measurements taken from the same Fe/Ge(001) system at RT show very small differences between the spectra obtained for a magnetic field applied in opposite directions. Furthermore, no magnetic hysteresis loop with remanence has been found at 150 K or at RT. This indicates a superparamagnetic behaviour of the Fe nanoclusters, which consist of sixteen Fe atoms, grown on the Ge(001) surface. Free Fe clusters of similar size were also shown to exhibit superparamagnetism [41].

Annealing the Fe nanoclusters on the Ge(001) surface at 420 K for 30 minutes leads to the formation of nanorow structures (or linear nanocluster arrays) elongated along the [1–10] direction of the Ge dimer rows (Fig. 6a). These were measured to be up to 15 nm in length (Fig. 6b). The temperature of 420 K has been chosen due to the observation that, for temperatures higher than 430 K interdiffusion of Fe and Ge seems to take place [42]. The Fe nanorow structures have a corrugation height of 1.1 Å, the same as the single nanoclusters prior to annealing. In the [110] direction, these nanostructures are approximately 9 Å wide. An individual nanorow consists of several Fe nanoclusters aligned along the direction of the Ge dimer rows [1–10]. The separation between the nanoclusters forming the nanorow is approximately 4 Å. This suggests that the cluster-cluster interaction is weaker than the interatomic interaction within a single cluster. This is in agreement with DFT results, showing that the cluster with the size of 9.0 Å × 9.0 Å formed by sixteen Fe atoms is the most energetically favourable on the Ge(001) surface.

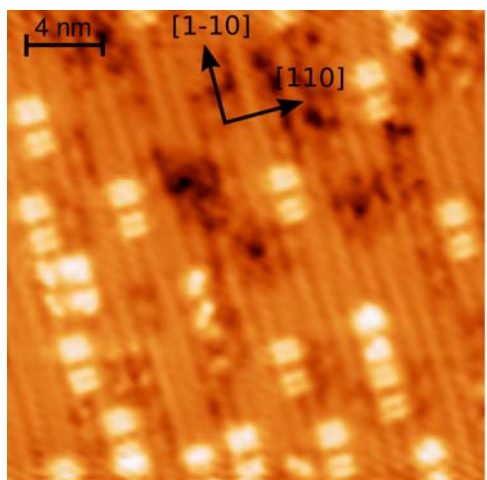 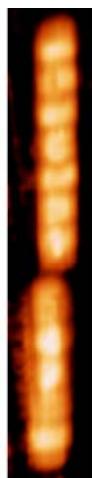 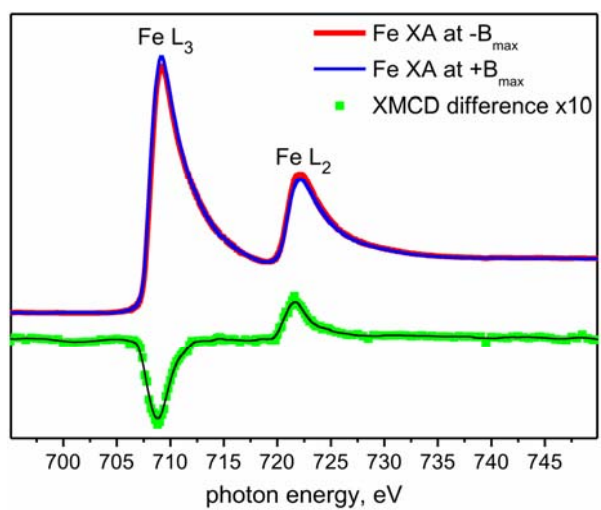

**Figure 6** Low-temperature STM images of 0.3 ML of Fe on the Ge(001) surface annealed at 420 K for 30 minutes to form nanorow structures (or linear nanocluster arrays). (a) $V_b = +1.5$ V, $I_t = 1.60$ nA, size 20.0 nm × 20.0 nm, 78 K. (b) $V_b = +1.5$ V, $I_t = 1.60$ nA, size 3.5 nm × 20.0 nm, 78 K.

**Figure 7** Fe 2p XA spectra measured at 150 K from 0.5 ML of Fe nanorow structures on the Ge(001) surface with the magnetic field of 0.05 T applied in two opposite directions ($-B_{max}$ and $+B_{max}$). The relative intensities of the spectra have been normalized to the same continuum jump at the photon energy of 745 eV, after subtraction of the background measured from a clean substrate. The XMCD spectrum is multiplied by a factor of 10 for clarity.

The XMCD measurements taken at RT and 150 K from the Fe nanorow structures grown on the Ge(001) surface also show that these structures exhibit superparamagnetic behaviour, similar to the separate Fe nanoclusters. The Fe 2p XA spectra measured from Fe nanorows at 150 K and the resulting XMCD spectrum are shown in Fig. 7. XMCD measurements for the Fe nanorow structures and the separate Fe nanoclusters show very small dichroism at RT and each exhibits a similar prominent dichroism at 150 K (see Figs. 5 and 7). In each case no magnetic hysteresis loop with remanence was observed, in comparison to the ferromagnetic response of larger clusters and nanorods [43, 44]. This suggests that an exchange interaction between the nanoclusters within the nanorow is not strong enough, or the nanorow size is still too small, to provide a ferromagnetic response.

## 4. Conclusions

The results obtained show that Ge(001) surface is a suitable template for the growth of homogeneous Fe nanoclusters. The size of nanoclusters depends on the temperature of the substrate. Deposition of 0.05–0.30 ML of Fe at RT and immediate cooling results in the formation of 4-atom Fe nanoclusters. Relaxation of the Fe/Ge(001) system at room temperature after deposition leads to the formation of nanoclusters with 16 Fe atoms. Annealing the nanoclusters at 420 K leads to the formation of nanorow structures with a width of 9 Å, due to cluster mobility at such temperature. The Fe nanoclusters and nanorow structures formed on the Ge(001) surface show a superparamagnetic behaviour as measured by XMCD.

**Acknowledgements**

This work was supported by Science Foundation Ireland (Principal Investigator grant number 06/IN.1/I91 and Research Frontiers Programme grant number 07/RFP/MASF185). The authors wish to thank Trinity College High Performance Cluster, funded by the Higher Education Authority under the Program for Research in Third Level Institutes, for the use of their computing facilities.